\documentclass[twocolumn]{aastex631}

\begin{document}

\title{Discovery of a Unique Close Quasar-DSFG Pair Linked by a [C II] Bridge at $z=5.63$}

\author[0000-0003-3307-7525]{Yongda Zhu}
\affiliation{Steward Observatory, University of Arizona, 933 North Cherry Avenue, Tucson, AZ 85721, USA; \href{mailto:yongdaz@arizona.edu}{\emph{yongdaz@arizona.edu}}}

% core data team

\author[0000-0002-5268-2221]{Tom J. L. C. Bakx} 
\affiliation{Department of Space, Earth, \& Environment, Chalmers University of Technology, Chalmersplatsen 4 412 96 Gothenburg, Sweden}

\author[0000-0002-2634-9169]{Ryota Ikeda}
\affiliation{Department of Astronomy, School of Science, SOKENDAI (The Graduate University for Advanced Studies), 2-21-1 Osawa, Mitaka, Tokyo 181-8588, Japan}
\affiliation{National Astronomical Observatory of Japan, 2-21-1 Osawa, Mitaka, Tokyo 181-8588, Japan}

\author[0000-0003-1937-0573]{Hideki Umehata}
\affiliation{Institute for Advanced Research, Nagoya University, Furocho, Chikusa, Nagoya 464-8602, Japan}
\affiliation{Department of Physics, Graduate School of Science, Nagoya University, Furocho, Chikusa, Nagoya 464-8602, Japan}

\author[0000-0003-2344-263X]{George D. Becker}
\affiliation{Department of Physics \& Astronomy,
    University of California, Riverside, CA 92521, USA}

\author[0000-0001-9420-7384]{Christopher Cain}
\affiliation{School of Earth and Space Exploration, Arizona State University,
Tempe, AZ 85287-6004, USA}

\author[0000-0002-6184-9097]{Jaclyn B. Champagne}
\affiliation{Steward Observatory, University of Arizona, 933 North Cherry Avenue, Tucson, AZ 85721, USA; \href{mailto:yongdaz@arizona.edu}{\emph{yongdaz@arizona.edu}}}

\author[0000-0003-3310-0131]{Xiaohui Fan}
\affiliation{Steward Observatory, University of Arizona, 933 North Cherry Avenue, Tucson, AZ 85721, USA; \href{mailto:yongdaz@arizona.edu}{\emph{yongdaz@arizona.edu}}}

\author[0000-0001-7440-8832]{Yoshinobu Fudamoto} 
\affiliation{Center for Frontier Science, Chiba University, 1-33 Yayoi-cho, Inage-ku, Chiba 263-8522, Japan}

\author[0000-0002-5768-738X]{Xiangyu Jin}
\affiliation{Steward Observatory, University of Arizona, 933 North Cherry Avenue, Tucson, AZ 85721, USA; \href{mailto:yongdaz@arizona.edu}{\emph{yongdaz@arizona.edu}}}

\author[0000-0002-5237-9433]{Hai-Xia Ma}
\affiliation{Division of Particle and Astrophysical Science, Nagoya University, Furo-cho, Chikusa-ku, Nagoya 464-8602, Japan}

\author[0000-0001-6561-9443]{Yang Sun} 
\affiliation{Steward Observatory, University of Arizona, 933 North Cherry Avenue, Tucson, AZ 85721, USA; \href{mailto:yongdaz@arizona.edu}{\emph{yongdaz@arizona.edu}}}

\author[0000-0001-8416-7673]{Tsutomu T.\ Takeuchi}
\affiliation{Division of Particle and Astrophysical Science, Nagoya University, Furo-cho, Chikusa-ku, Nagoya 464-8602, Japan}
\affiliation{The Research Center for Statistical Machine Learning, the Institute of Statistical Mathematics, 10-3 Midori-cho, Tachikawa, Tokyo 190-8562, Japan}

\author[0000-0003-0747-1780]{Wei Leong Tee}
\affiliation{Steward Observatory, University of Arizona, 933 North Cherry Avenue, Tucson, AZ 85721, USA; \href{mailto:yongdaz@arizona.edu}{\emph{yongdaz@arizona.edu}}}

\begin{abstract}
We report the discovery of a unique quasar-dusty star-forming galaxy (DSFG) system at $z = 5.63$, consisting of the bright quasar J1133+1603 ($M_{\rm UV} = -27.42$) and its compact, dust-obscured companion, J1133c. ALMA observations reveal a prominent [\ion{C}{2}] bridge connecting the quasar and DSFG, indicating ongoing interaction at a projected separation of 1.8\arcsec\ ($\sim$10 proper kpc). J1133c exhibits unusually bright and broad [\ion{C}{2}] emission ($L_{\rm [CII]} > 10^{43}\, \rm erg\, s^{-1}$, $\rm FWHM > 500\, \rm km\,s^{-1}$), with a [\ion{C}{2}] luminosity five times that of the quasar, suggesting intense star formation or potential AGN activity. The inferred star formation rate from [\ion{C}{2}] is approximately $10^3 M_\odot \rm yr^{-1}$. The remarkable properties of this pair strongly suggest that galaxy interactions may simultaneously trigger both starburst and quasar activity, driving rapid evolution in the early universe.
\end{abstract}

\section{Introduction}
Mergers are considered significant triggers of quasar activity and star formation, with up to 50\% of luminous high-redshift quasars linked to major mergers \citep[e.g.,][]{di_matteo_energy_2005,hopkins_unified_2006,decarli_rapidly_2017}. High-redshift quasar-dusty star-forming galaxy (DSFG), quasar-quasar, and quasar-submillimeter galaxy (SMG) pairs \citep[e.g.,][]{lee_first_2019, izumi_merging_2024, decarli_alma_2019-1} are rare but essential for studying galaxy-quasar co-evolution during the reionization epoch. Such systems typically reside in dense environments where interactions profoundly impact both star formation and black hole growth, as seen in SMGs and active galactic nuclei (AGNs) \citep[e.g.,][]{alexander_rapid_2005,alexander_weighing_2008,umehata_alma_2015}.

The [\ion{C}{2}] 158$\mu$m line can trace star formation and gas dynamics, offering insights into interstellar medium properties of AGNs and star-forming galaxies \citep[e.g.,][]{stacey_158_1991,lagache_cii_2018}. The quasar-DSFG pair reported here, connected by a [\ion{C}{2}] bridge, provides a unique opportunity to explore the effects of such interactions on the growth of massive galaxies. Similar [\ion{C}{2}] bridges have been noted in other high-redshift systems, indicating active gas flows between interacting galaxies \citep[e.g.,][]{ginolfi_alpine-alma_2020,umehata_alma_2021}. This work uses the {\tt Planck18} cosmology \citep[][]{planck_collaboration_planck_2020} for luminosity calculations.

\section{ALMA Observations}
\begin{figure*}[!ht]
    \centering
    \includegraphics[width=1\linewidth]{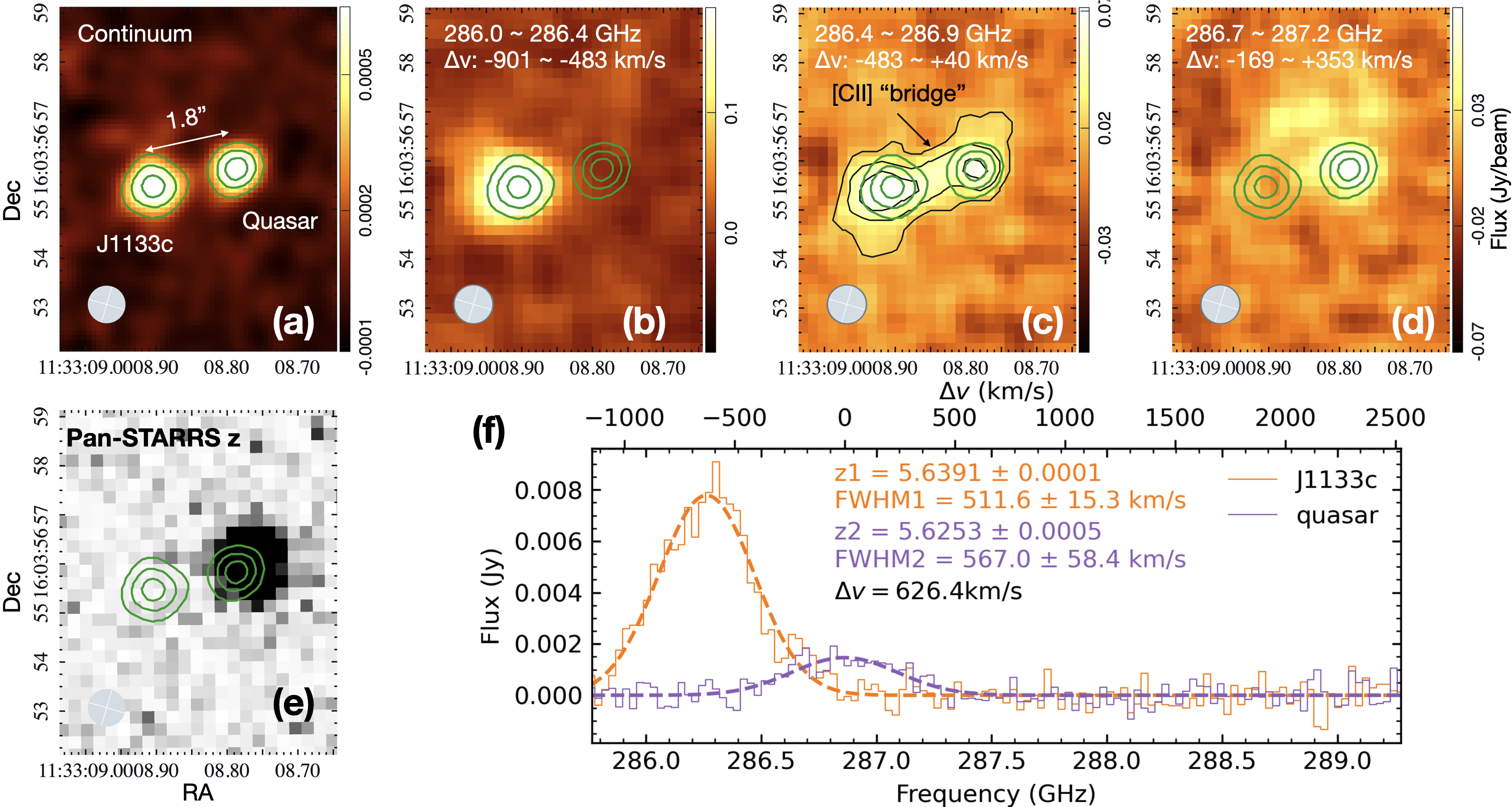}
    \caption{(a) ALMA continuum image of the quasar-DSFG system. (b, c, d) ALMA [\ion{C}{2}] emission stacked over $286.0\sim286.4$ GHz, $286.4\sim286.9$ GHz, and $286.7\sim287.2$ GHz, showing J1133c, the [\ion{C}{2}] bridge, and the quasar host, with labeled velocity offsets ($\Delta v$). (e) Pan-STARRS z-band image for reference. Green contours indicate $5\sigma$, $9\sigma$, and $13\sigma$ continuum detections, with black contours in (c) showing $1\sigma$ and $2\sigma$ [\ion{C}{2}] bridge detections. The ALMA beam size is shown in gray. (f) [\ion{C}{2}] spectra of J1133c and the quasar with Gaussian fits, labeling central redshifts, FWHMs, and the velocity offset.}
    \label{fig:1}
\end{figure*}
In Cycle 9 (2022.1.00662.S; PI: Zhu), we conducted ALMA Band 7 observations targeting 21 quasars at $z \sim 5.5$ identified in \citet{yang_discovery_2017,yang_filling_2019}, aiming to determine their precise systemic redshifts ($\Delta z \sim 10^{-4}$) through [\ion{C}{2}] 158$\mu$m line emission (details in \citealp{zhu_probing_2023}). Each observation included two overlapping spectral windows centered on the estimated redshift for [\ion{C}{2}] line coverage, along with two additional spectral windows for dust continuum measurements. The observations utilized the C43-(1, 2, and 3) configurations, achieving an angular resolution of  $\sim 1''$ across the whole sample.

The data were calibrated and reduced using the CASA pipeline (version 6.4.1.12; \citealp{mcmullin_casa_2007,casa_team_casa_2022}). We created data cubes and imaged the [\ion{C}{2}] emission following  \citet{eilers_detecting_2020}. Continuum subtraction was performed using {\tt uvcontsub}, and imaging was executed with {\tt tclean} using Briggs weighting with a robust parameter of 2 (natural weighting) to enhance sensitivity. The mean RMS noise level across our dataset was $\sim 0.25\ {\rm mJy\,beam^{-1}}$ per 30 MHz bin.

\section{Results}
The quasar J1133+1603 (RA=11:33:08.78, Dec=+16:03:55.79) was observed with an integration time of 1118.88 seconds, yielding a continuum sensitivity of 0.0293 ${\rm mJy\,beam^{-1}}$ and an angular resolution of 0.632\arcsec. The companion, J1133c (RA=11:33:08.90, Dec=+16:03:55.47), lies southeast of the quasar at a projected separation of 1.8\arcsec\ ($\sim$10 kpc; Figure \ref{fig:1}a). Both the quasar and companion appear relatively compact in the continuum, with FWHMs of $\sim 0.84\arcsec$ and $\sim 0.7\arcsec$ along the major axis, respectively. The continuum flux density is $9.57\times 10^{-4}$ Jy for the quasar and $1.07\times 10^{-3}$ Jy for J1133c. Despite J1133c being as bright as the quasar in the continuum and brighter in [\ion{C}{2}] (see below), it is undetected in Pan-STARRS $z$-band imaging, likely due to the survey's shallow depth ($m_{z} = 20.9$; Figure \ref{fig:1}e).

We analyzed the [\ion{C}{2}] emission, finding a luminosity of $2.88\times 10^{42} \,\rm erg\,s^{-1}$ for the quasar and $1.43\times10^{43}\,\rm erg\,s^{-1}$ for J1133c, with the latter being 4.97 times higher. The star formation rate for J1133c, estimated from [\ion{C}{2}], is $\log \rm [ SFR / (M_\odot \, yr^{-1}) ]\sim 3$ \citep[e.g.,][]{lagache_cii_2018}. Figures \ref{fig:1} (b) and (d) show [\ion{C}{2}] emission from J1133c and the quasar, respectively. Notably, J1133c exhibits strong [\ion{C}{2}] emission, while Figure \ref{fig:1} (c) shows a [\ion{C}{2}] bridge linking the two, suggesting active interaction. Extended emission north of the quasar (panel d) further supports this scenario.

Panel (f) displays the [\ion{C}{2}] spectra of both objects with best-fit Gaussians. The quasar’s [\ion{C}{2}] redshift is $5.6253\pm0.0005$ with FWHM $567.0\pm 58.4\, \rm km\,s^{-1}$, while J1133c has $z=5.6391\pm 0.0001$ and FWHM $511.6\pm 15.3 \,\rm km\,s^{-1}$. The velocity offset between them is $\Delta v=626.4 \,\rm km\,s^{-1}$. These redshifts may be affected by kinematics within the system, complicating separation measurements along the line of sight.

J1133c’s large velocity width ($\rm FWHM > 500\, km\,s^{-1}$) is broader than the average [\ion{C}{2}] FWHM of $\sim 300$ km\,s$^{-1}$ in typical ALPINE galaxies \citep{bethermin_alpine-alma_2020}, marking it as a notable outlier. This suggests strong dynamical processes or feedback within J1133c, potentially due to interaction with the quasar.

\section{Discussion}
The discovery of this quasar-DSFG system at $z=5.63$, linked by a [\ion{C}{2}] bridge, provides a valuable case for studying galaxy and supermassive black hole co-evolution during reionization. The relatively compact morphology of J1133c in both its dust continuum (FWHM $\sim$ 0.84\arcsec) and [\ion{C}{2}] emission (FWHM $\sim$ 1.01\arcsec) suggests a small star-forming region or gas stripped due to interaction. This compactness is particularly noteworthy compared to other known quasar companions at $z > 5.5$ \citep[e.g.,][]{decarli_rapidly_2017} and raises the possibility of an obscured AGN within J1133c. The intense star formation and/or potential AGN activity in J1133c could result from an ongoing merger.

This system displays a smaller distance between the quasar and DSFG compared to other well-studied pairs, such as BRI 1202-0725 at $z=4.7$, with a separation of 3.8\arcsec\ (25 kpc) \citep[e.g.,][]{lee_first_2019,2020ApJ...902...37D}. The [\ion{C}{2}] bridge observed between J1133+1603 and J1133c suggests a direct interaction, potentially indicating a merger process. Similar gas bridges have been observed in other high-redshift systems, such as PSOJ167$-$13 and J1306+0356 at $z > 6$, where [\ion{C}{2}] and dust continuum emission connect both galaxies \citep[][see also \citealp{2022MNRAS.517L..11S} for a CO bridge at $z\sim 2.6$]{neeleman_resolved_2019}. Extended [\ion{C}{2}] emission seen to the north of the quasar, spanning several kpc, further supports complex gas dynamics from this interaction. These characteristics highlight the role of mergers and gas flows in early galaxy formation and black hole growth, making this system a great case for exploring galaxy evolution and quasar activity in dense environments.

\section*{acknowledgments}
YZ acknowledges support from the NIRCam Science Team contract to the University of Arizona, NAS5-02015. YZ and GDB were supported by the NSF through award SOSPADA-029 from the NRAO, and grant AST-1751404. CC acknowledges support from the Beus Center for Cosmic Foundations.
JBC acknowledges funding from the JWST Arizona/Steward Postdoc in Early galaxies and Reionization (JASPER) Scholar contract at the University of Arizona.
This paper makes use of the following ALMA data: ADS/JAO.ALMA\allowbreak\#2022.1.00662.S. ALMA is a partnership of ESO (representing its member states), NSF (USA) and NINS (Japan), together with NRC (Canada), MOST and ASIAA (Taiwan), and KASI (Republic of Korea), in cooperation with the Republic of Chile. The Joint ALMA Observatory is operated by ESO, AUI/NRAO and NAOJ. The National Radio Astronomy Observatory is a facility of the National Science Foundation operated under cooperative agreement by Associated Universities, Inc.

\vspace{5mm}
\facilities{ALMA}

\software{
    {\tt CARTA} \citep{comrie_carta_2021},
    {\tt CASA} \citep{mcmullin_casa_2007,casa_team_casa_2022}
}

%\bibliography{references}{}
%\bibliographystyle{aasjournal}

\end{document}